**Room-temperature ferroelectric nematic liquid crystal showing a large and divergent density**

Charles Parton-Barr[1], Helen F. Gleeson[1], Richard J. Mandle[1,2,*]
[1] School of Physics and Astronomy, University of Leeds, Leeds, LS2 9JT, UK
[2] School of Chemistry, University of Leeds, Leeds, LS2 9JT, UK
[*]corresponding author email: r.mandle@leeds.ac.uk

**Abstract**
The ferroelectric nematic phase ($N_F$) is a recently discovered phase of matter in which the orientational order of the conventional nematic liquid crystal state is augmented with polar order. Atomistic simulations suggest that the polar $N_F$ phase would be denser than conventional nematics owing to contributions from polar order. Using an oscillating U-tube densitometer, we obtain detailed temperature-dependent density values for a selection of conventional liquid crystals with excellent agreement with earlier reports. Having demonstrated the validity of our method, we then record density as a function of temperature for M5, a novel room-temperature ferroelectric nematic material. We present the first experimental density data for a $N_F$ material as well as density data for a nematic that has not previously been reported. We find that the room-temperature $N_F$ material shows a large (>1.3 g cm$^3$) density at all temperatures studied, with an increase in density at phase transitions. The magnitude of the increase for the intermediate splay- ferroelectric nematic ($N_X$-$N_F$) transition is an order of magnitude smaller than the isotropic-nematic (I-N) transition. Present results may be typical of ferroelectric nematic materials, potentially guiding material development, and is especially relevant for informing ongoing studies into this emerging class of materials.

**Introduction**
The nematic (N) phase is the simplest liquid crystalline state, consisting of molecules or particles that have long-range orientational order but lack translational order. Despite this orientational order, the bulk nematic phase is apolar as there are an equal number of molecules oriented parallel and antiparallel.

In the recently discovered ferroelectric nematic ($N_F$) phase[1-6], the orientational order of the conventional nematic state is augmented by so-called polar order which arises due to parallel alignment of molecular electric dipole moments. This parallel alignment distinguishes it from the nematic phase, and virtually all other fluid states of matter. Depictions of nematic and polar nematic phases are given in Fig. 1.

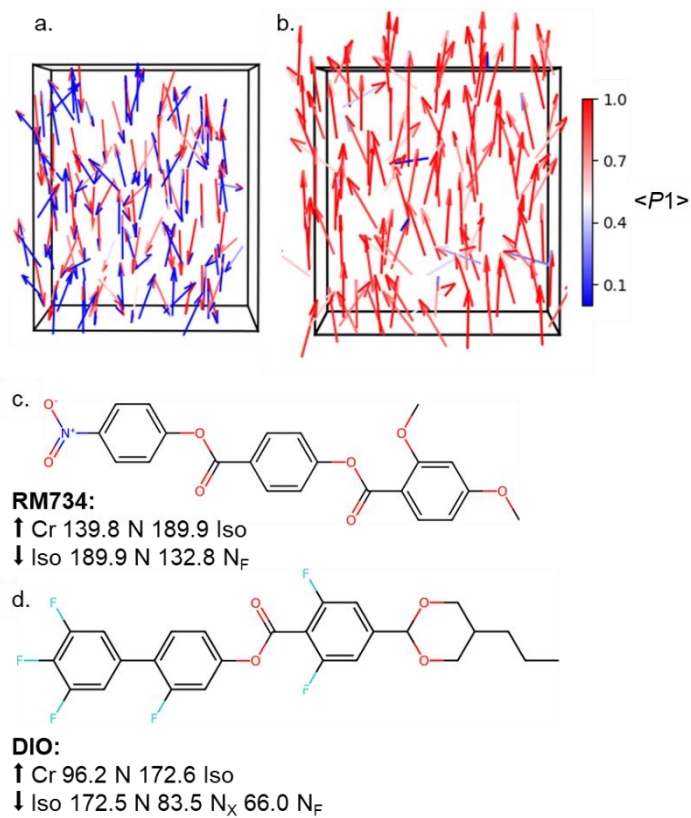

**Figure 1:** Depictions of the orientations of the molecular electric dipole moments in the (a) nematic and (b) ferroelectric nematic phases; the arrows are color-coded to represent the contribution to the bulk value of the polar order parameter, $<P1>$ which is 0.08 for (a) and 0.86 for (b), respectively, while $<P2>$ is 0.66 and 0.66 for (a) and (b), respectively. In both cases the director is oriented perfectly with the box height (z-axis). The molecular structures of the first reported $N_F$ materials; (c) RM734 [1, 2], and (d) DIO [3] along with their transition temperatures (T, °C) on heating (↑) and cooling (↓)

Having been the subject of some speculation [7-9], the $N_F$ phase has been now experimentally realised in several classes of materials, having originally been observed in RM734 and DIO, (Fig. 1 (c, d)), both reported in 2017. In recent years the number of $N_F$ materials has been increased significantly as derivatives of RM734 and DIO [10-14] have been developed, including those exhibiting a direct I-$N_F$ transition [15] .

In RM734 [1, 2] the N-$N_F$ transition is accompanied by a softening of the $K_1$ splay constant, and a growth of ferroelectric ordering [4, 6]. DIO [3] behaves slightly differently, with an intermediate phase between the N and $N_F$; crucially however, DIO was also shown to possess these same characteristic macroscopic domains of ferroelectric ordering in the low temperature nematic phase [5]. When studied by DSC, there is a small enthalpy associated with the transition from nematic to polar nematic phase(s). Dielectric measurements have reported large dielectric permittivity values on the order of $10^4$ for RM734[16] and for DIO [3, 17], although the validity of these has been questioned [18]. Polarisation investigations have found values of spontaneous polarisation for RM734 ($6\mu C/cm^2$) [5, 19] and DIO (2.5-6 $\mu C/cm^2$)[3, 17] which are comparable with polar columnar ($5.8\mu C/cm^2$) [20] and bent core liquid crystals (LCs) ($0.5-0.8\mu C/cm^2$) [21]. The measured $N_F$ spontaneous polarisation values sit between that of the chiral smectic C, DOBAMBAC, (42.0 $nC/cm^2$) [22] and the inorganic material PbSc$_{0.5}$Ta$_{0.5}$O$_3$ (30.0 $\mu C/cm^2$) [23].

Our investigations were driven by the predictions of atomistic molecular dynamics (MD) simulations. MD simulations can reproduce the polar nematic ordering and calculate mean densities from atomic scale interactions [24]. Simulations predicted a high mass density, $\rho$ of 1.3 $g/cm^3$ [24] in RM734. Considering the oft-used assumption that $\rho$ in liquid crystals $\approx$ 1 $/cm^3$ , known to be an approximation for some LCs and is unjustified for many others, this raises a potential defining property of N$_F$ materials.

As LCs undergo a first-order phase transition, they exhibit a density change that can be discontinuous [25]. The temperature dependence of density, $\rho(T)$ in liquid crystals can be described by a linear thermal expansion of the specific volume, $v_{sp}$ as described in equation (1) and (2)

$$v_{sp}(T) = 1/\rho(T) \qquad (1)$$

$$v_{sp}(T) = v_{sp}^* + \alpha T \qquad (2)$$

where $v_{sp}^*$ is the specific volume at 0°C, $\alpha$ is an empirical expansion coefficient describing how the material behaves with temperature changes, and $T$ is the temperature.

The density of LCs has previously been measured by several methods including the capillary tube technique [24], the weight change of a submerged glass cylinder [26] and a dilatometer measuring the height of a mercury interface [27]. These methods all suffer from a difficulty in obtaining the precise temperature control that detailed study over phase transitions demands. Moreover, the lack of experimental density data for all but a handful of common LCs has led to the incorrect (*vide infra*) assumption about $\rho$ being propagated.

Density data were available for 5CB [28-32], 8CB [33-35], and (NCS)PCH6 [36]. A comprehensive list of the data and their properties are given in section 4 of the supplementary information. The 5CB dataset [29] was chosen for comparison with high resolution data (between 5°C and 0.2°C) over a large temperature range. The available data for 8CB consisted of either sparse measurements [34] or data taken over a temperature range near a single transition [33]. The (NCS)PCH6 literature data [36] were used out of necessity as no other data were available. We estimate the agreement of our data through Bland-Altman [37] type plots (SI section S4). We find the limits of agreement through equation (3) where $\bar{d}$ is the mean difference between the literature and $\sigma_d$ is the standard deviation of the differences and judge the agreement through their vicinity to $\bar{d}$.

$$\bar{d} = 1.96\sigma_d \qquad (3)$$

The largest difference between the agreement comes from a systematic differences between the methods in the capillary tube investigation of 5CB [31] and 8CB [35] where a dilatometer is used. These methods have a systematic difference of approximately 1% and 3% respectively compared to 0.02% [29] and 0.04% [34] of the density meters. This is calculated from the average of the differences used in the Bland-Altman plots (Fig. S9-S17 in SI).

In this article we report an unusually large density (>$1.3 gcm^3$) for a ferroelectric nematic liquid crystal, M5 (Merck Electronics KGaA). A density increase resulting from polar ordering on transition to an $N_F$ state is also shown. We benchmark our results against a set of standard liquid crystals, which we find to be in excellent agreement with the available literature data.

**Experimental**
**Density Measurements**

Density measurements were performed using an Anton-Paar DMA 4100M densitometer which operates on the oscillating U-tube principle (Fig. 2). A U-shaped borosilicate glass tube is filled with ≈ $1 ml$ material. Samples were loaded using a SGE precision syringe. A syringe heater was used for samples with above-ambient melting points; the syringe outlet was interfaced with the DMA 4100M inlet port using a short section of PTFE tubing with appropriate PEEK/ETFE microfluidic connectors (purchased from Darwin Microfluidics). The presence of bubbles within the tube was assessed through visual inspection. This experimental method requires a large amount of material; however, it can be largely recovered on the completion of the measurements. Recovered material was subjected to purification with a Teledyne Combiflash NextGen 300+ Flash chromatography system and filtration over a 0.2 micron PTFE filter.

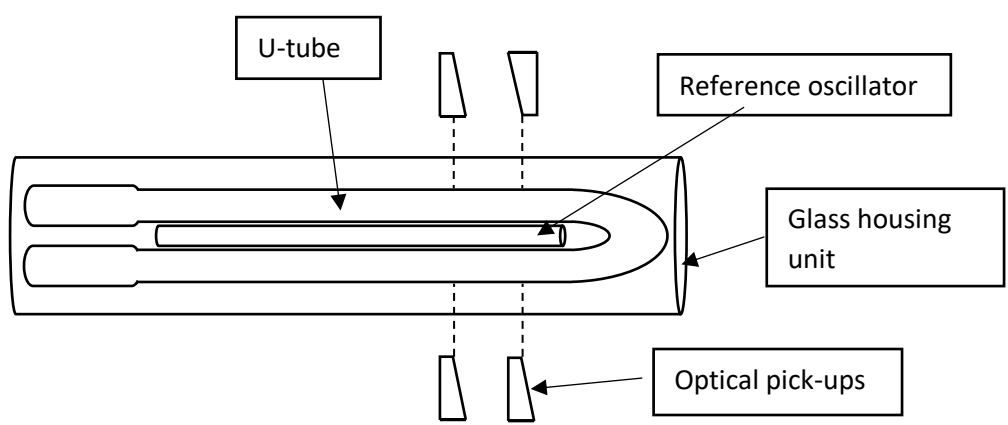

**Figure 2**. Schematic diagram of the oscillating u-tube setup used to measure density. The optical pickups determine the LC sample's characteristic frequency which corresponds to a density value.

Once filled, the tube is excited to oscillate at its characteristic frequency in the direction of the normal to the tube. The density is calculated from equation (4)

$$\rho = A \cdot Q^2 \cdot f_1 - B \cdot f_2 \qquad (4)$$

with $Q$ being the quotient of the characteristic frequencies of the measuring tube and a reference tube. $A$ and $B$ are constants specific to the instrument and are calculated by calibrating against substances with a precisely known value for density. $f_1$ and $f_2$ are corrections factors. The $f_1$ and $f_2$ correction factors possess contributions from the viscosity of the sample. An increase in viscosity has an increased damping effect on the characteristic frequency of the oscillating tube, resulting in a greater magnitude of corrections needed. A valid measurement for this machine is three successive measurements that do not deviate beyond $0.0001 g/cm^3$. The repeatability of the temperature during measurement is $0.02°C$.

**Materials**

The materials selected were M5 (Merck Electronics KGaA), 4-cyano-4-pentylbiphenyl (5CB, Fluorochem), 4-cyano-4-octylbiphenyl (8CB, Fluorochem), 1-(trans-4-hexylcyclohexyl)-4-isothiocyanatobenzene ((NCS)PCH6), Sigma-Aldrich) and trans-4-(trans-4'-n-propylcyclohexyl)-cyclohexyl-3,4,5-trifluorobenzene (CCU-3-F, prepared according to ref [38]). Both M5 and commercial materials were used as received. Figure 3 presents the transition temperatures for the materials investigated in this work, and chemical structures in the case of single-component materials (Fig. 3a-d).

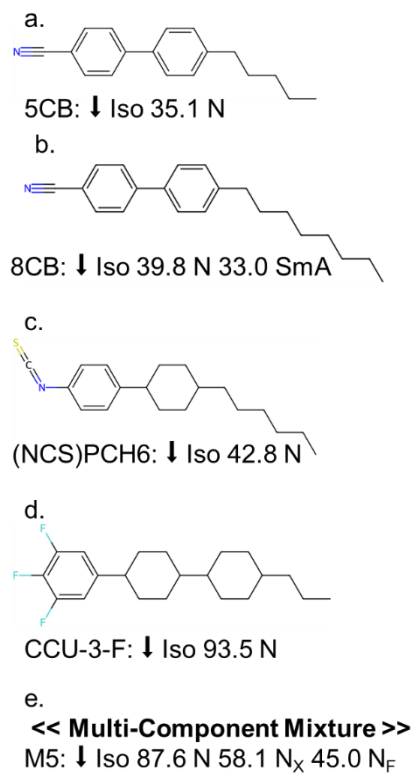

a.
5CB: ↓ Iso 35.1 N

b.
8CB: ↓ Iso 39.8 N 33.0 SmA

c.
(NCS)PCH6: ↓ Iso 42.8 N

d.
CCU-3-F: ↓ Iso 93.5 N

e.
<< Multi-Component Mixture >>
M5: ↓ Iso 87.6 N 58.1 $N_X$ 45.0 $N_F$

**Figure 3.** Chemical structures of the materials used in this work. Transition temperatures (T, °C) on cooling (↓) were determined from density measurements and are in good agreement with DSC data (SI section S1). For single component materials (a-d) molecular structures are given.

## Results
### Apolar Liquid Crystals

Our initial investigations focused on a set of relatively well-known materials, for some of which density data was available. We selected 5CB as it is an ambient temperature nematic; 8CB as it exhibits a transition between two LC phases (smectic A (SmA), a layered structure and N) and has a high viscosity at ambient temperature; (NCS)PCH6 as a non-nitrile room-temperature nematic LC; CCU-3-F as a fluorinated LC with an above-ambient melting point which enabled us to refine our technique for handling materials with elevated melting points. The temperature-dependent densities for the selected materials are given in Fig. 4.

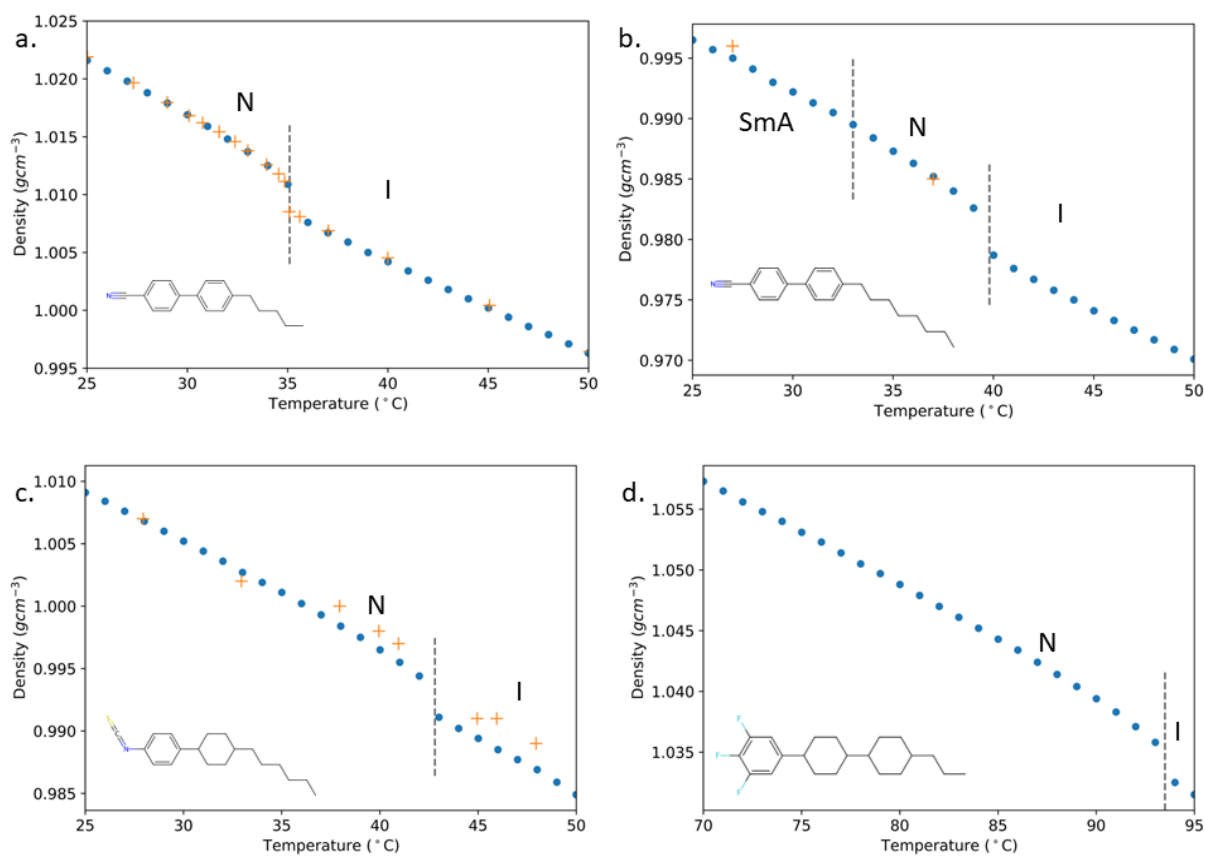

**Figure 4.** Densities of the liquid crystal materials at 1°C intervals. Dashed lines denote transition temperatures as defined by the temperature of the largest density gradient. The density increases at the phase transition are readily observable for the I-N transitions, but more subtle for the N-SmA transition. Comparison is made with literature data where possible (orange plus points) for (a) 5CB [29], (b) 8CB [34] and (c) (NCS)PCH6

[36]. (d) The CCU-3-F figure is representative of the scarcity of LC density data.

The measurement runs for all materials began in the isotropic phase and were cooled through the phase range into their lowest temperature phase above crystallisation. The present results compare very favourably with available literature data for 5CB, 8CB and (NCS)PCH6. In all cases, the density of the LC materials is around 1 $g/cm^3$. The I-N transitions for each material can be clearly observed as discontinuous increase in $\rho$. At the N-SmA transition we do not find a discernible increase in density, however it can be seen in Fig. S8

**Polar Liquid Crystal**

Having shown our ability to reproduce and expand upon literature densities of common liquid crystals, we next performed density measurements on the room-temperature $N_F$ material M5, a multi-component mixture produced by Merck Electronics KGaA. A cursory inspection of the data in Fig. 5 reveals that, even in the isotropic liquid, the density of M5 is remarkably high, being ~ 30% larger than 5CB and of the same magnitude as dichloromethane (1.33 g cm$^3$ at 25 °C).

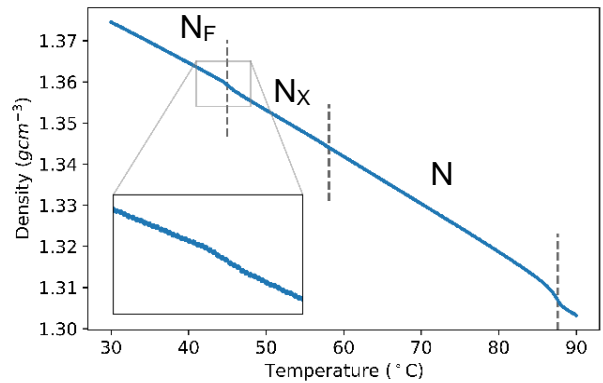

**Figure 5.** Temperature-dependent density behaviour for the $N_F$ material M5 across its phase range. The initial increase with first nematic ordering can be seen on cooling from the isotropic phase at 87.6°C similar to the other selected materials. There is no discernible change in density at the N-$N_X$ phase transition. There is a small increase at 45°C that can be attributed to the effects of polar ordering. The high average density of M5 should also be noted. Inset: $N_X$-$N_F$ transition

In M5 there is a constant decrease in the expansion coefficients as it is cooled down through its phase range (Table 1). This can also be observed in the rate of density change in the linear regimes of Fig. 5. As with the other materials investigated, an increase in the density can be seen at $T_{IN}$ and this feature is also clear at $T_{N_X N_F}$ though the $T_{NN_x}$ density change appears to be continuous. Fig. 5 presents data obtained at an increased temperature resolution (0.15°C compared to the 1°C given in Fig. 4). The inset facilitates a more detailed look at the behaviour over the $N_X$-$N_F$ transition where the density increase can be seen.

**Features of Temperature Dependent Density Behaviour**

| Material | $\alpha 10^4$ $(cm^3 g^{-1} K^{-1})$ | | | | |
|---|---|---|---|---|---|
|  | $N_F$ | $N_X$ | SmA | N | I |
| M5 | 5.30 | 6.08 | - | 6.57 | 7.09 |
| 5CB | - | - | - | 9.18 | 7.96 |
| 8CB | - | - | 8.16 | 11.5 | 8.76 |
| CCU-3-F | - | - | - | 7.81 | 9.00 |
| (NCS)PCH6 | - | - | - | 8.08 | 8.80 |

**Table 1.** Expansion coefficients, $\alpha$, of $v_{sp}$ calculated for each material in each phase. These parameters are analogous to the thermal expansion coefficient describing how a material's volume changes with temperature. The parameters are obtained through linear fits to $v_{sp}(T)$ within each specific LC phase.

The linear relationship between density and temperature that exists within a single phase can be explored through equation (2). From this, $v_{sp}(T)$ is separated into discrete phases using the transition temperatures in Fig. 4 as their temperature ranges. The gradient of a linear fit of $v_{sp}(T)$ for each phase (see also Fig. S18 in SI) is taken to be the expansion coefficient. Expansion coefficients for the selected materials can be found in table 2. The expansion coefficients are similar, although we do not find an explicit relationship between coefficient and LC phase type as the coefficient can be seen to either increase (5CB, 8CB) or decrease (CCU-3-F, (NCS)PCH6) through the I-N transition.

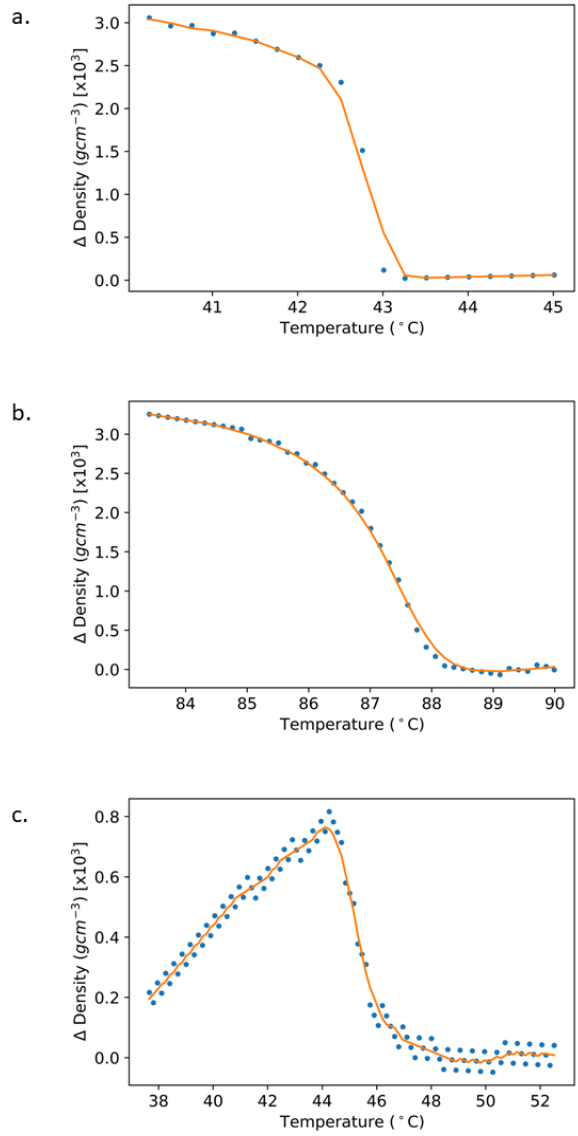

**Figure 6**. The density deviation ($\Delta\rho = \rho(T) - \rho^{highT}(T)$) through the I-N transition (a, b) of (NCS)PCH6 and M5 respectively. The behaviour in (a, b) is the result of a conventional nematic ordering taking place. C) shows how the density changes in M5 as a result of polar ordering. Data is presented as scatter points whereas the solid line is a Savitzky-Golay filter [39] smoothing intended to show the trend of $\Delta\rho$.

|  | $\Delta\rho 10^3$ $(g/cm^3)$ | | | |
|---|---|---|---|---|
| **Material** | I-N | N-$N_X$ | $N_X$-$N_F$ | N-SmA |
| M5 | 3.1 | 0.05 | 0.75 | - |
| 5CB | 2.2 | - | - | - |
| 8CB | 2.5 | - | - | 0.20 |
| CCU-3-F | 2.0 | - | - | - |
| (NCS)PCH6 | 2.3 | - | - | - |

**Table 2**. The magnitude of the density increases of each transition through the non-linear $\rho(T)$ region. The I-N transitions for all materials are of a similar magnitude as the same sort of ordering takes place. M5's $N_X$-$N_F$ transition is an order of magnitude

larger than its N-N$_X$ transition and three times the size of the N-SmA transition of 8CB.

Table 2 gives the density increases for the selected materials. A linear fit of the higher temperature phase $\rho^{highT}$ was extrapolated into the lower temperature phase and used to calculate the density deviation of the lower temperature phase as it cools. A representative example of this process is given in Fig. S6. The density deviation at (NCS)PCH6's I-N transition can be seen in Fig. 6a. We find Δρ of the I-N transitions for all materials studied to be of the same order of magnitude, presumably a consequence the fact the same type of molecular reorganisation takes place for each material at $T_{IN}$. For 8CB we find the N-SmA transition is ten times smaller than its I-N transition indicating that the change in molecular packing at the smectic transition is not as a drastic as a nematic one.

M5's increase in density from its I-N and N$_X$-N$_F$ phase transitions can be seen in Fig. 6 (b, c). The N$_X$-N$_F$ transition is accompanied by a much smaller density change than that associated with the I-N transition, approximately $0.00075 g/cm^3$ vs $0.0031$ $g/cm^3$ (table 2). The size of this I-N density change is the largest of the investigated materials. The density deviation of the N$_X$-N$_F$ transition has been calculated at more than three times larger than that of the N-SmA transition. This suggests a significant structural reorganisation taking place through polar ordering when compared to that of the positional ordering of the N-SmA transition. It is interesting to note that M5 does not follow the loose positive correlation between density increase and transitional enthalpy (supplementary table 2) perhaps due to its composition as a mixture.

**Conclusion**

Density influences a myriad of fundamental liquid crystal properties including the birefringence [40] and order parameters [41]. We have shown that the density of the N$_F$ material M5 is significantly larger than conventional liquid crystals, a finding that is in keeping with that measured experimentally in the solid state atomistic MD simulations [24]. Here, we find that the N$_F$ material M5 possesses a much higher density (~$1.3 g/cm^3$) than that of conventional nematic systems where the density is assumed (and found) to be ~$1 g/cm^3$. We also observe experimentally an increase in density on entering the N$_F$ phase which is comparable to that seen in silico [24]

For classical apolar liquid crystals it is customary to assume a density of 1 $g/cm^3$ if required for measurement or analysis. In the case of N$_F$ materials, we suggest that, in the absence of data for a specific material, a value of 1.35 $g/cm^3$ be used.

**Acknowledgements**

The authors thank Merck Electronics KGaA for gifting the sample of M5 used in this work, and for their continued partnership and support. HFG thanks EPSRC for funding via grant number EP/V054724/1. RJM thanks UKRI for funding *via* a Future Leaders Fellowship, grant number MR/W006391/1, and UoL for funding a PhD studentship for CPB. The authors all thank Diana Nikolova of the University of Leeds for providing DSC data for M5.

# Supplementary Information

# Room-temperature ferroelectric nematic liquid crystal showing a large and divergent density

Charles Parton-Barr[1], Helen F. Gleeson[1], Richard J. Mandle[1,2,*]
[1] School of Physics and Astronomy, University of Leeds, Leeds, LS2 9JT, UK
[2] School of Chemistry, University of Leeds, Leeds, LS2 9JT, UK
*corresponding author email: r.mandle@leeds.ac.uk

Contents:

**SECTION S1 - DSC SCANS**

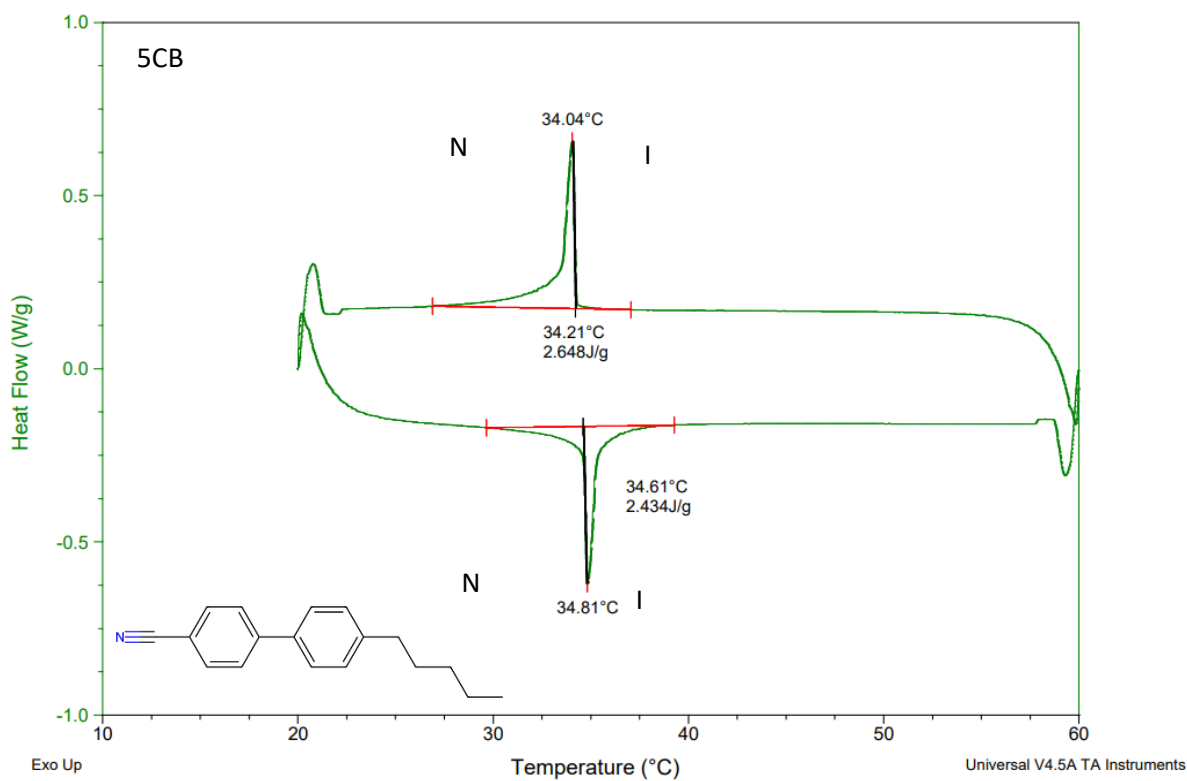

**Figure S1**: DSC Thermogram (endo up, exo down) for 5CB recorded at (10 °C min$^{-1}$). The phase state of liquid crystal through the measurement run is labelled.

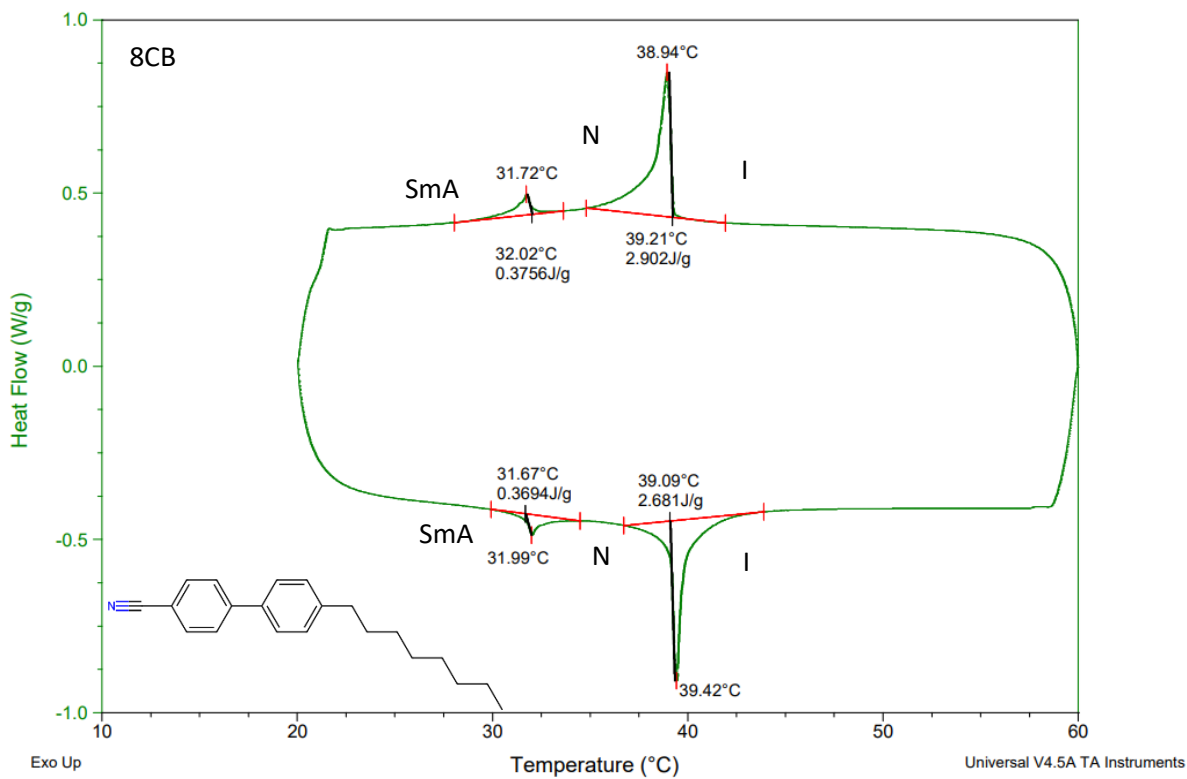

**Figure S2**: DSC Thermogram (endo up, exo down) for 8CB recorded at (10 °C min$^{-1}$). The phase state of liquid crystal through the measurement run is labelled.

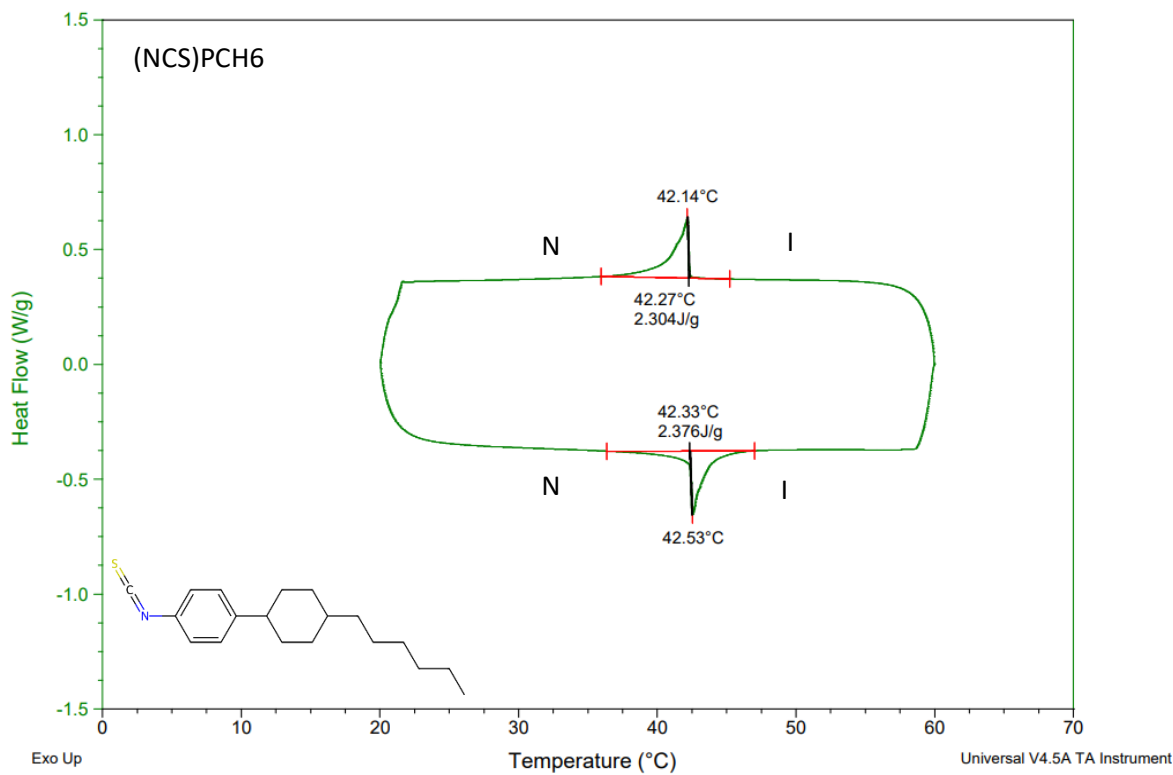

**Figure S3**: DSC Thermogram (endo up, exo down) for 8CB recorded at (10 °C min$^{-1}$). The phase state of liquid crystal through the measurement run is labelled

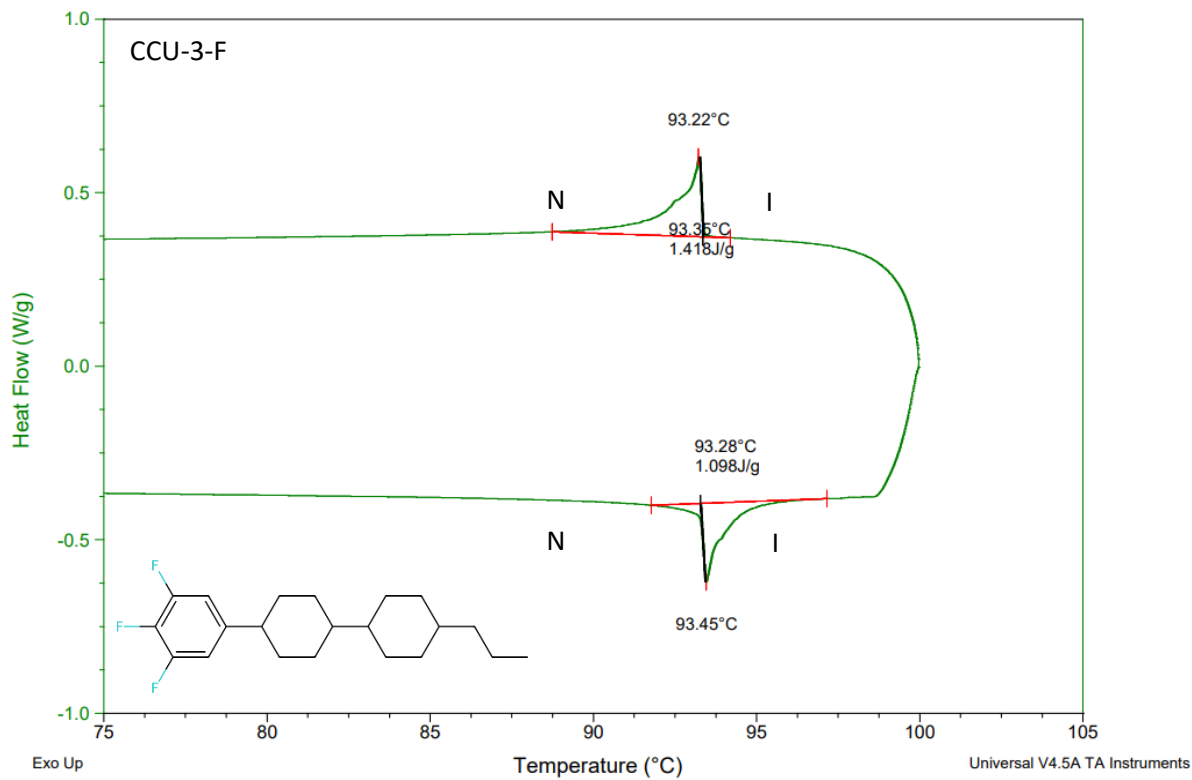

**Figure S4**. DSC Thermogram (endo up, exo down) for CCU-3-F recorded at (10 °C min$^{-1}$). The phase state of liquid crystal through the measurement run is labelled.

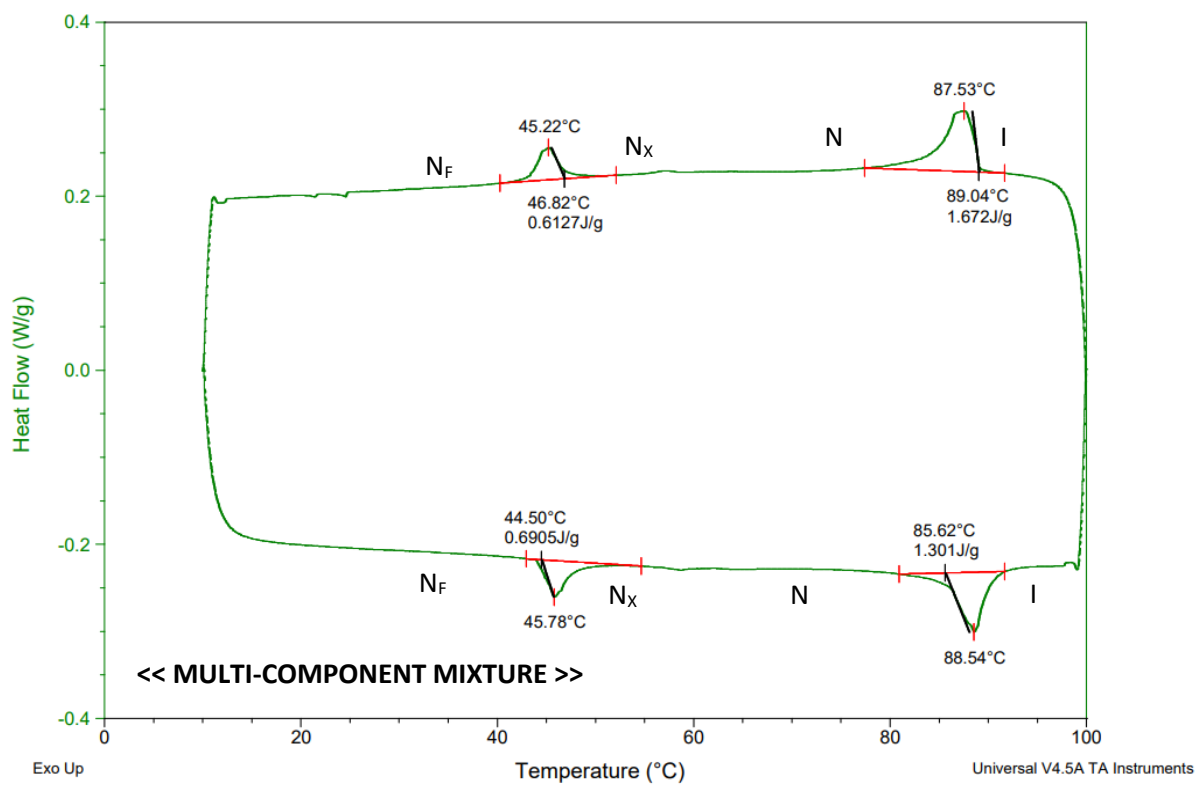

**Figure S5**. DSC Thermogram (endo up, exo down) for M5 recorded at (10 °C min$^{-1}$). The phase state of liquid crystal through the measurement run is labelled.

## SECTION S2 - NORMALISED DENSITY DEVIATION EXAMPLE

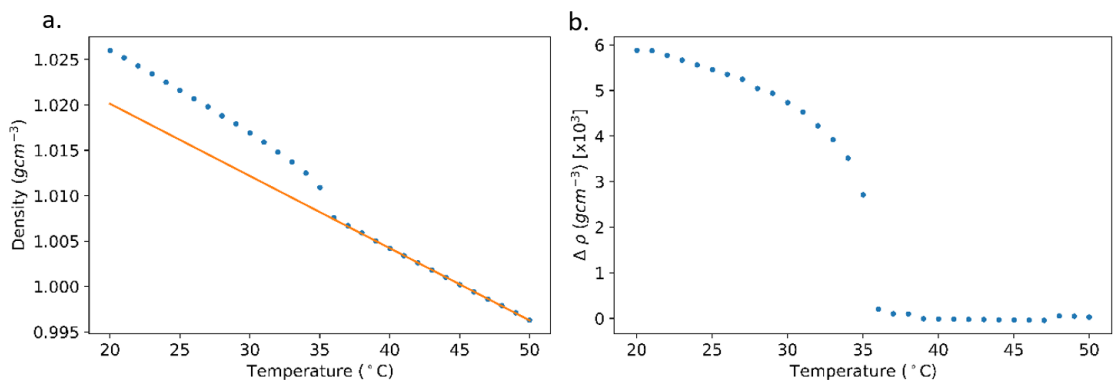

**Figure S6**. An example to illustrate how the deviation of the density at a transition, $\Delta\rho(T)$ is examined. Densities from a measurement run of 5CB are given in (a) where a linear fit of the density values in the higher temperature phase, $\rho^{highT}(T)$ is also shown. The resulting deviation $\Delta\rho(T) = \rho(T) - \rho^{highT}$ is then shown in (b).

## SECTION S3 - VISCOSITY CORRECTIONS

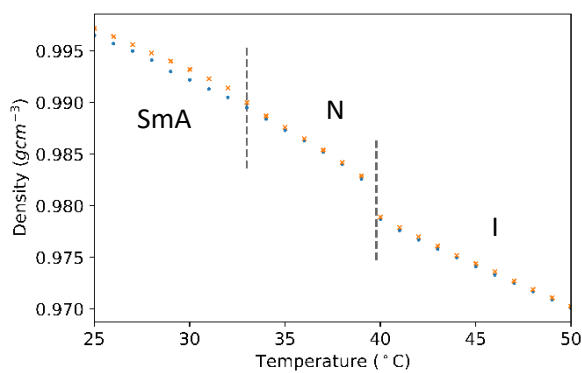

**Figure S7**. The density, $\rho(T)$ of 8CB. An increased discrepancy between the viscosity corrected (circles) and viscosity non-corrected (crosses) can be seen as 8CB is cooled through SmA-N

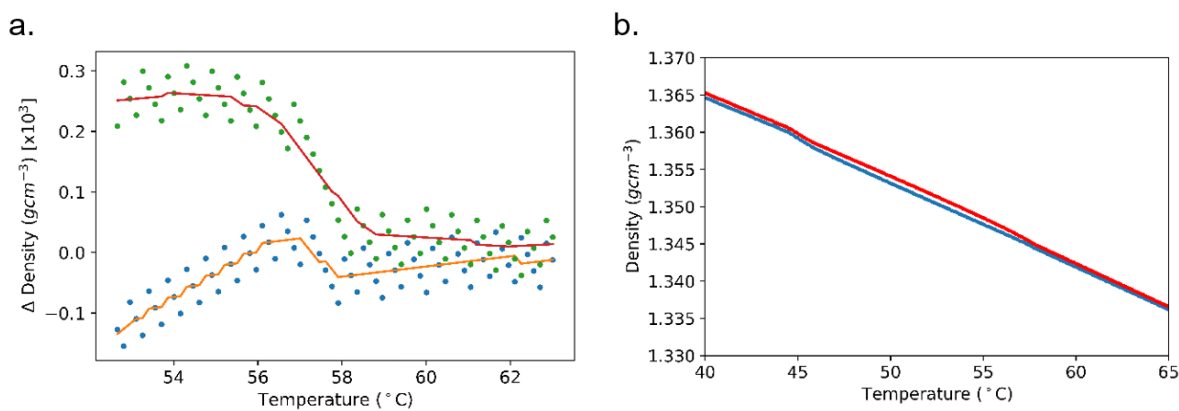

**Figure S8**. These figures show discounting viscosity corrections for M5 can affect the density values over the (a) N-$N_X$ transition and (b) a temperature range containing N-$N_X$-$N_F$ phases. In (a) the density data without viscosity corrections are shown as the green dots with the red visual guideline. (b) gives the data without viscosity corrections as the red points.

The N-$N_X$ transition has the smallest $\Delta\rho$ of M5's transitions and does not appear to show any distinct increase in Fig 6. Fig. S8 aims to replicate the visualisation of the I-N and $N_X$-$N_F$ transitions for N-$N_X$. The density measurements without viscosity corrections produce a greater deviation at the $N_X$-$N_F$ which is shown in Fig. 8a. This same procedure can be applied to Fig. 4b to produce a similar plot of the SmA-N transition (Fig. S7). Fig. 8b shows a discrepancy that grows at N-$N_X$ and is maintained through it the $N_X$ phase. It is interesting to note that it decreases at $N_F$-$N_X$ before growing again on cooling.

## SECTION S4 - LITERATURE DATA COMPARISONS

| LC | Source | Method | Temp range °C | Resolution °C | Agreement |
|---|---|---|---|---|---|
| 5CB | Oweimreen [1] | DMA60 oscillating density meter. | 35.0-35.6 | 0.07>x>0.04 | Fig. S9 |
| 5CB | Shimada [2] | DA-640 oscillating density meter | 19.0-50.0 | 5.0>x>0.2 | Fig. S10 |
| 5CB | Zeller [3] | N/A | -10.0 – 50.0 | N/A | Fig. S11 |
| 5CB | Tintaru [4] | Capillary method | 26.5-48.0 | 2.0>x>0.5 | Fig. S12 |
| 5CB | Deschamps [5] | Pycnometer | 20.0-65.0 | 13.0 > x > 5.0 | Fig. S13 |
| 8CB | Oweimreen [6] | DMA60 oscillating density meter | 33.56-33.88 | 0.05 > x > 0.02 | Fig. S14 |
| 8CB | Dunmur [7] | DMA 02 oscillating density meter | 27.0-37.0 | 10.0 | Fig. S15 |
| 8CB | Leadbetter [8] | Dilatometer | 27.0-50.0 | N/A | Fig. S16 |
| (NCS)PCH6 | Baran [9] | Pycnometer | 23.0-53.0 | 1.0>x>5.0 | Fig. S17 |

**Supplementary Table 1.** The available literature data for the selected liquid crystals. The table states the method used to obtain the data, the range over which the data was taken and the minimum and maximum resolution of the data.

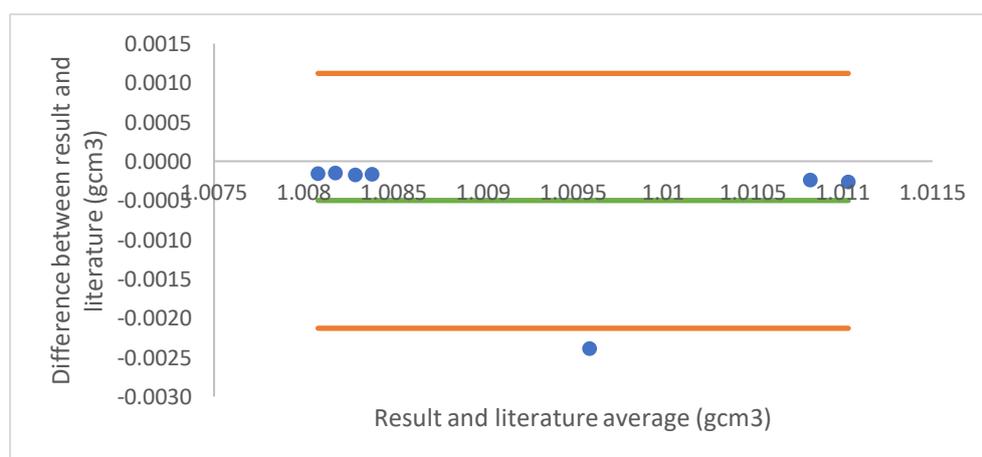

**Figure S9.** The means of the literature value and measured value are plotted against the difference between the two values (blue points). The orange lines denote the upper and lower bounds of two standard deviations. The green line gives the mean difference between literature and measured values.

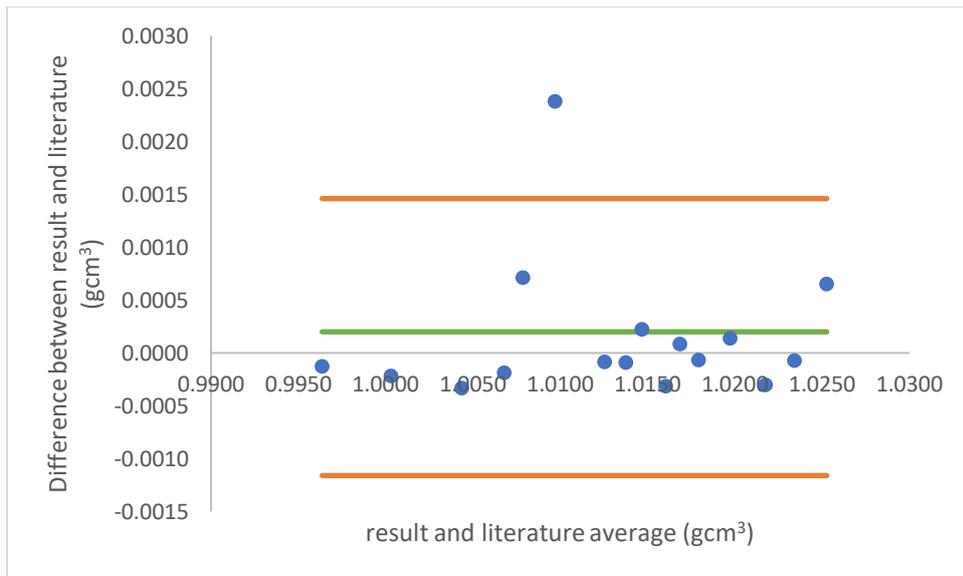

**Figure S10**. The means of the literature value and measured value are plotted against the difference between the two values (blue points). The orange lines denote the upper and lower bounds of two standard deviations. The green line gives the mean difference between literature and measured values.

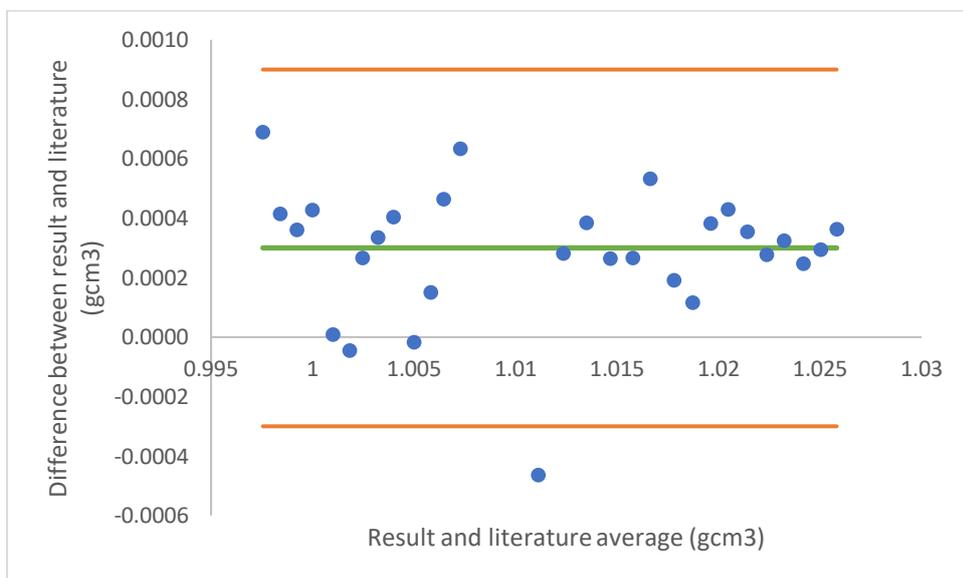

**Figure S11**. The means of the literature value and measured value are plotted against the difference between the two values (blue points). The orange lines denote the upper and lower bounds of two standard deviations. The green line gives the mean difference between literature and measured values.

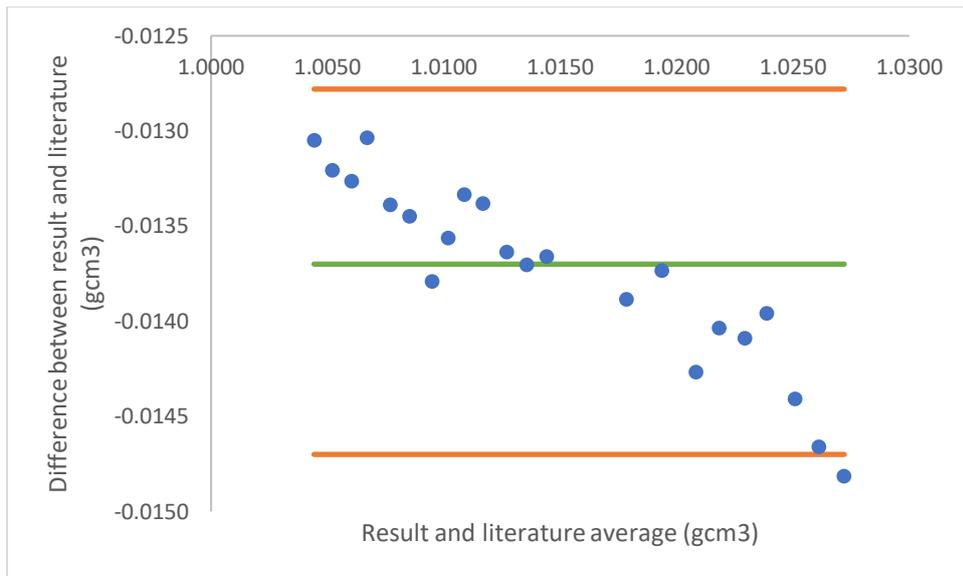

**Figure S12**. The means of the literature value and measured value are plotted against the difference between the two values (blue points). The orange lines denote the upper and lower bounds of two standard deviations. The green line gives the mean difference between literature and measured values.

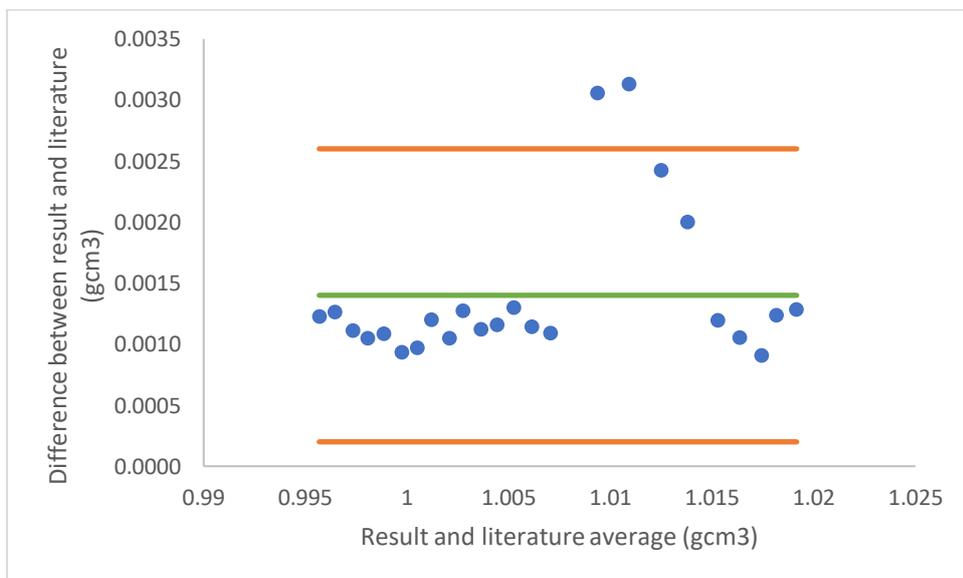

**Figure S13**. The means of the literature value and measured value are plotted against the difference between the two values (blue points). The orange lines denote the upper and lower bounds of two standard deviations. The green line gives the mean difference between literature and measured values.

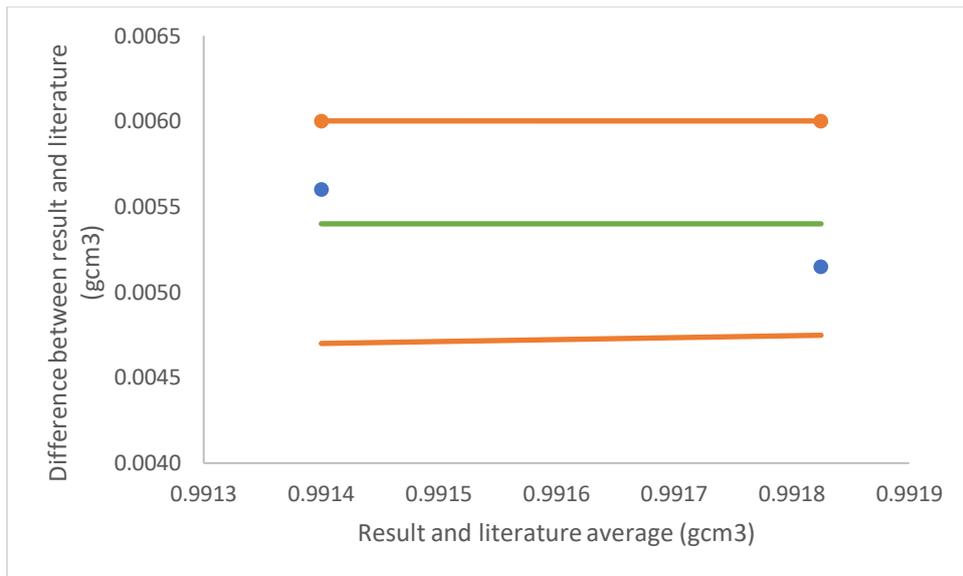

**Figure S14**. The means of the literature value and measured value are plotted against the difference between the two values (blue points). The orange lines denote the upper and lower bounds of two standard deviations. The green line gives the mean difference between literature and measured values.

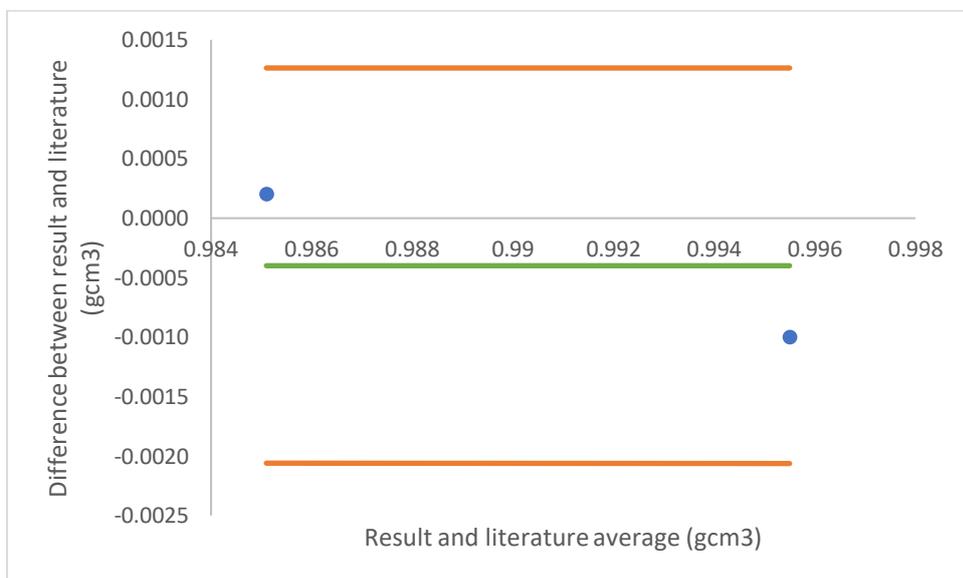

**Figure S15**. The means of the literature value and measured value are plotted against the difference between the two values (blue points). The orange lines denote the upper and lower bounds of two standard deviations. The green line gives the mean difference between literature and measured values.

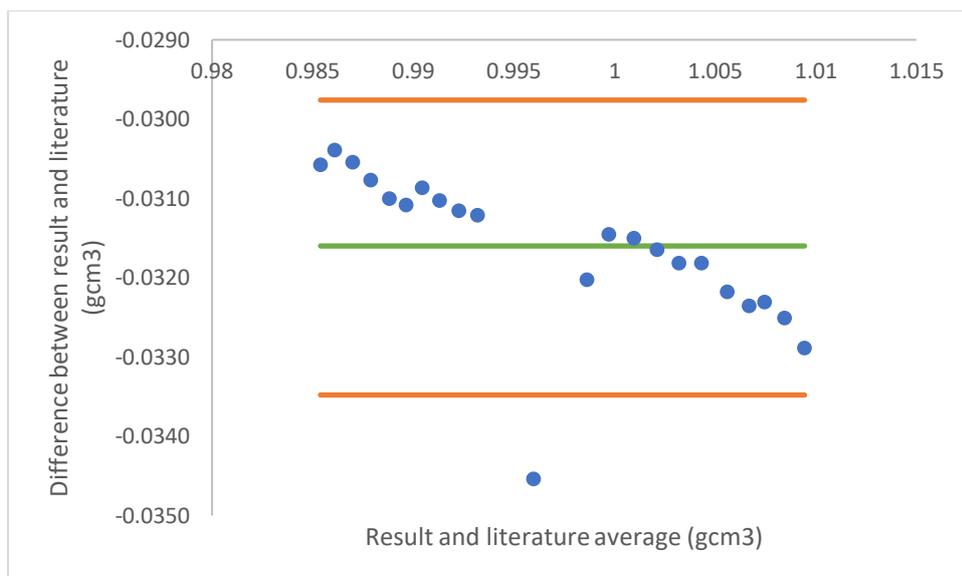

**Figure S16**. The means of the literature value and measured value are plotted against the difference between the two values (blue points). The orange lines denote the upper and lower bounds of two standard deviations. The green line gives the mean difference between literature and measured values.

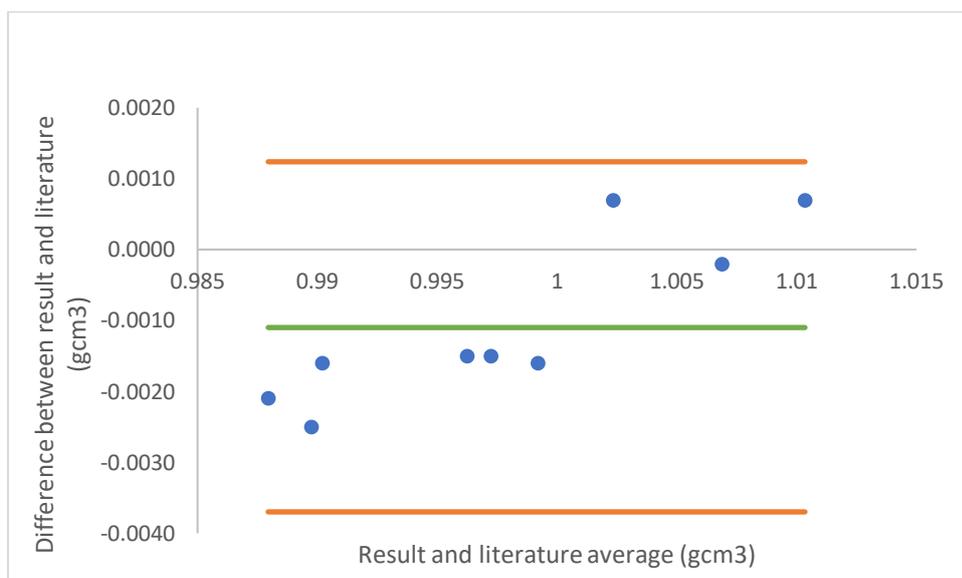

**Figure S17**. The means of the literature value and measured value are plotted against the difference between the two values (blue points). The orange lines denote the upper and lower bounds of two standard deviations. The green line gives the mean difference between literature and measured values.

**SECTION S5 – TEMPERATURE DEPENDENT DENSITY BEHAVIOUR**

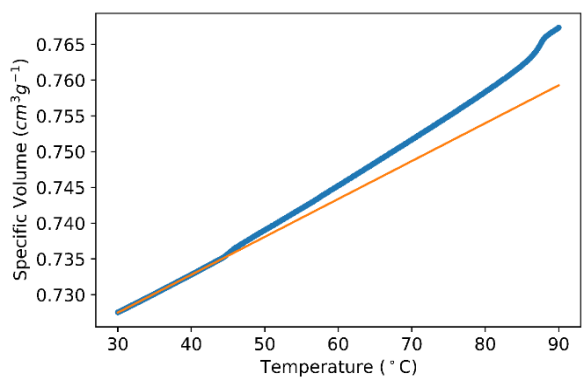

**Figure S18**. The specific volume (1/density) of M5. The orange line represents the linear fit of the density data that is used to calculate a value for the expansion coefficient, $\alpha$

The calculated expansion coefficients, $\alpha$, vary through the phases for every material (table 1). The density data provides different values for the expansion coefficient of N and $N_X$ in M5 as well as the SmA and N phases in 8CB. These correspond to phases in which their transitions do not produce clear deviations. This provides an alternative method for identifying the phase transition through analysis of the specific volume's gradient.

Using our DSC data (SI section 1), a loose correlation between the transition enthalpies and the density increases, $\Delta\rho$ can be observed. There will also be a contribution to the density increase from the expansion coefficient with the larger values adding a larger increase to the volume over the transition. M5 differs from the selected materials as it is a mixture. Possibly due to this the transitional enthalpy and density increase of M5 does not follow the correlations of the single component materials. Its larger N-I density increase (Fig 6. (b), supplementary table 2) is a result of the transition density increase taking place over a larger temperature range ($\approx$1.5K vs. $\approx$0.5K) than the conventional material. However, its transitional enthalpy is closest to that of CCU-3-F and smaller than the other materials.

| Material | Transition Enthalpies (J/g) | $\Delta\rho 10^3 \ (g/cm^3)$ |
|---|---|---|
| M5 | 1.424 | 3.1 |
| 5CB | 2.648 | 2.2 |
| 8CB | 2.902 | 2.5 |
| CCU-3-F | 1.418 | 1.9 |
| (NCS)PCH6 | 2.304 | 2.3 |

**Supplementary Table 2.** A comparison between the transition enthalpies and density deviations of the I-N transitions for the LC materials. There appears to be a loose relationship between the size of the transition enthalpies and the magnitude of the density deviation for the selected materials. This does not hold true for the multi-component M5.

We have noted that considering the contribution of the viscosity to the corrections made by the density meter can allow for the graphical observation of transitions. The viscosity behaviour of liquid crystals is highly complex and outside the scope of this work. However, it is an interesting parallel that the viscosity correction occurs in both 8CB and M5. If we assume that there is some increase in the viscosity of M5 at $N_X$-N (like that of SmA-N of 8CB) then the discrepancy between the corrected and uncorrected density values can potentially be understood. This is a logical direction to be explored in future work.